\documentclass[sigconf]{acmart}

\pdfoutput=1
\usepackage[misc]{ifsym}
\usepackage{balance}
\usepackage{mathtools}
\usepackage{bm}
\usepackage{multirow}
\usepackage{subfigure}
\usepackage{tabularray}
\usepackage{makecell}

\AtBeginDocument{%
  }

\copyrightyear{2024}
\acmYear{2024}
\setcopyright{acmlicensed}
\acmConference[CIKM '24]{Proceedings of the 33rd ACM International Conference on Information and Knowledge Management}{October 21--25, 2024}{Boise, ID, USA}
\acmBooktitle{Proceedings of the 33rd ACM International Conference on Information and Knowledge Management (CIKM '24), October 21--25, 2024, Boise, ID, USA}
\acmDOI{10.1145/3627673.3679663}
\acmISBN{979-8-4007-0436-9/24/10}

\begin{document}

\title{Aligning Explanations for Recommendation with Rating and Feature via Maximizing Mutual Information}

\author{Yurou Zhao}
\email{zhaoyurou@ruc.edu.cn}
\affiliation{%
  \institution{Gaoling School of Artificial Intelligence, Renmin University of China}
  \city{Beijing}
  \country{China}
}

\author{Yiding Sun}
\email{emanual20.sun@foxmail.com}
\affiliation{%
  \institution{Gaoling School of Artificial Intelligence, Renmin University of China}
  \city{Beijing}
  \country{China}
}

\author{Ruidong Han}
\email{hanruidong@meituan.com}
\affiliation{%
  \institution{Meituan}
  \city{Beijing}
  \country{China}
}

\author{Fei Jiang}
\email{jiangfei05@meituan.com}
\affiliation{%
  \institution{Meituan}
  \city{Beijing}
  \country{China}
}

\author{Lu Guan}
\email{guanlu02@meituan.com}
\affiliation{%
  \institution{Meituan}
  \city{Beijing}
  \country{China}
}

\author{Xiang Li}
\email{lixiang245@meituan.com}
\affiliation{%
  \institution{Meituan}
  \city{Beijing}
  \country{China}
}

\author{Wei Lin}
\email{linwei31@meituan.com}
\affiliation{%
  \institution{Meituan}
  \city{Beijing}
  \country{China}
}

\author{Weizhi Ma}
\email{mawz@tsinghua.edu.cn}
\affiliation{%
  \institution{Tsinghua University}
  \city{Beijing}
  \country{China}
}

% \author{Jiaxin Mao\textsuperscript{*}}\thanks{*Corresponding author}
\author{Jiaxin Mao}
\authornote{Corresponding author}
\email{maojiaxin@gmail.com}
\affiliation{%
  \institution{Gaoling School of Artificial Intelligence, Renmin University of China}
  \city{Beijing}
  \country{China}
}
\renewcommand{\shortauthors}{Yurou Zhao et al.}


\begin{abstract}
Providing natural language-based explanations to justify recommendations helps to improve users' satisfaction and gain users’ trust. However, as current explanation generation methods are commonly trained with an objective to mimic existing user reviews, the generated explanations are often not aligned with the predicted ratings or some important features of the recommended items, and thus, are suboptimal in helping users make informed decision on the recommendation platform. To tackle this problem, we propose a flexible model-agnostic method named MMI (\textbf{M}aximizing \textbf{M}utual \textbf{I}nformation) framework to enhance the alignment between the generated natural language explanations and the predicted rating/important item features. Specifically, we propose to use mutual information (MI) as a measure for the alignment and train a neural MI estimator. Then, we treat a well-trained explanation generation model as the backbone model and further fine-tune it through reinforcement learning with guidance from the  MI estimator, which rewards a generated explanation that is more aligned with the predicted rating or a pre-defined feature of the recommended item. Experiments on three datasets demonstrate
that our MMI framework can boost different backbone models, enabling them to outperform existing baselines in terms of alignment with predicted ratings and item features. Additionally, user studies verify that MI-enhanced explanations indeed facilitate users' decisions and are favorable compared with other baselines due to their better alignment properties.
% implication: 
\end{abstract}






\maketitle

\section{Introduction}
 Generating natural language-based explanations for recommendation has gained wide attention in recent years \cite{pepler,peter,NETE,saer}. A series of studies have shown the potential benefits of providing explanations on recommendation platforms, such as increasing the acceptance ratio of recommendations \cite{efm,ex3,spotify} and users' satisfaction and trust \cite{chen2022measuring,exrec-survey,7goals}. To generate fluent and personalized explanations, most studies leverage existing user reviews as ground truth and mainly use the Maximum Likelihood Estimation (MLE) approach to train models capable of generating explanations that resemble user reviews. Although such practices are promising to generate high-quality text in terms of traditional text generation metrics (e.g., BLEU, ROUGE), they struggle to meet the users' additional requirement of explanation on the recommendation platform, such as trust and effectiveness \cite{7goals}. Serving as auxiliary information for the recommended item, the explanation is expected to facilitate users to make more informed decisions on the platform \cite{7goals,conflicting-goal,effectiveness}. To achieve that, a qualified explanation is expected to embody the following two properties: 
 \\
\textbf{Alignment with Predicted Rating}: The explanation ought to support the recommender's predicted rating, as it could potentially assist users in comprehending the rationale behind a specific recommendation. A conspicuous discrepancy between the sentiment conveyed in the explanation and the predicted rating (e.g., an explanation stating ``the decor is nice and the staff is friendly'' for an item rated 2/5 star) will mislead the users into making wrong decisions and foster their skepticism towards the recommendation platform.
 \\
 \textbf{Alignment with Item Features}:  The explanation needs to include relevant and highly specific information about a certain feature or aspect of the recommended item. Generic explanations like "good food and good service" are not informative enough for users to gain detailed knowledge about the recommended item, and thus, do not assist them in deciding whether to accept or reject the corresponding recommendation.
% \begin{table*}
% \centering
% \caption{Case Study of Commonly Adopted User Reviews Dataset Yelp for Explanation Generation}
% \label{tab:bad_user_review}
% \scalebox{0.9}{
% \begin{tblr}{
%   row{1} = {c},
%   row{2} = {c},
%   cell{3}{2} = {c},
%   cell{3}{3} = {c},
%   cell{3}{4} = {c},
%   cell{3}{5} = {c},
%   cell{3}{6} = {c},
%   vline{2} = {-}{},
%   hline{1-2,4} = {-}{},
% }
%        & User                   & Item             & {Predicted \\ Rating }& {Explanation perfectly \\ reconstructing  the review    } & {Explanation  aligning with \\predicted rating/item feature}        \\
% Case A & 7hnwrH20r6IiJOAlNOPgcw & Bonefish Grill             & 2.0    & the staff was very good                                   & {the sauce was \textbf{bland} and \\ the texture was \textbf{too thick}} \\
% Case B & 9lipnr92ElS6EqDxLnQEiA & Hot Wok Chinese Restaurant & 3.0    & it was okay .                                             & the \textbf{prices }are very reasonable                             
% \end{tblr}
% }
% \end{table*}

 Unfortunately, solely mimicking user reviews is not sufficient to fulfill the above two goals. 
The reasons are: 1) Noise in user reviews: User reviews are widely used as a proxy for training explanation generation models~\cite{nrt,att2seq,peter,pepler,NETE,apref2seq,saer,dualpc}. However, we notice that user reviews often contain non-explanatory content such as purely subjective narratives and extremely generic comments. Such contents neither describe the details about the product nor reflect the reason why the user gives the product a positive/negative rating. As a result, models trained on user reviews may be influenced by this noise and thus fail to align well with the rating or feature.
2) The average nature of MLE training objective: 
% Admittedly, the irrelevant or unqualified content in user reviews can be filtered out by designing a strict and sophisticated data-cleaning pipeline. However, 
Even if current models are trained on an ideal dataset, the commonly adopted MLE still hinders them from obtaining better alignment. Since MLE intrinsically favors high-frequency, generic phrases over more specific, contextually relevant explanations, it will make the trained model generate sentences that are rather generic and lack diversity. This tendency will contribute to the poor alignment of explanation with rating and feature. On one hand,  the reviews in the dataset are mostly sentimentally positive, which makes the existing explanation generator constantly generate positive sentences for all recommendations even if the actual predicted rating is low. On the other hand, the distribution of item features in the dataset is uneven. As a result, the explanation generator is prone to mention common yet less specific features (e.g. food/service for the Yelp dataset as these two features are dominant in the dataset ) for each item.

To intuitively illustrate the limitations of exploiting user reviews as explanations, Table \ref{tab:bad_user_review} presents two cases comparing the explanations with strong alignment properties, and the explanations completely recover the corresponding user reviews which are deemed perfect in most previous works as they achieve high value of NLG metrics on the user review dataset. 
In Case A, the user is predicted to give the item a low rating, suggesting dissatisfaction. However, the explanation that shares high similarity with the review  contradicts the negative sentiment  of the predicted rating. On the contrary, the explanation better aligned with the predicted rating sheds light on disappointing issues with the item, such as the sauce being bland and the texture being too thick. Such explanation is more likely to help the user understand the reason behind the recommender's predicted rating and thus potentially 
 increase user-perceived transparency of the recommender. 
In Case B, the explanation copying the corresponding review presents a generic statement, which lacks specific details regarding the description of the item features. Meanwhile, the explanation better aligned with the item feature highlights a particular item feature (prices) that the user may find valuable when deciding whether to accept or reject the recommendation. This kind of explanation provides users with a more informative understanding of the recommended item, enabling them to efficiently make decisions on the recommendation platform.
\begin{table}
\centering
\caption{Illustration of the importance of strengthening explanation's alignment properties}
\label{tab:bad_user_review}
\scalebox{0.8}{
\begin{tblr}{
  row{1} = {c},
  row{2} = {c},
  cell{3}{2} = {c},
  cell{3}{3} = {c},
  cell{3}{4} = {c},
  vline{2} = {-}{},
  hline{1-2,4} = {-}{},
}
       & {Predicted \\ Rating} & {Explanation perfectly \\ reconstructing  the review} & {Explanation  aligning with \\predicted rating/item feature}           \\
Case A & 2.0                   & the staff was very good                               & {the sauce was \textbf{bland} and\\the texture was \textbf{too thick}} \\
Case B & 3.0                   & it was okay .                                         & the \textbf{prices }are very reasonable                                
\end{tblr}}
\end{table}

To address the above limitations, we propose a  model-agnostic Maximizing Mutual Information (MMI) framework for strengthening the alignment ability of current explanation generation models. As mutual information is a principal measure of the mutual dependence between the two variables, we utilize it to measure to what extent the explanation is aligned with the predicted rating or item feature. The MMI framework features: 1) a neural MI estimator \cite{mine} to estimate the alignment between text-based explanations and the predicted rating/item features; 2) an RL-based fine-tuning process that treats an existing MLE-trained explanation generation model as the backbone and fine-tunes it with the MI-based reward output by the MI estimator. To avoid potential reward hacking and maintain the backbone model's ability to mimic user reviews, we also integrate KL and Entropy reward as regularizers to enable the fine-tuned generator to strike a good balance between the ability of alignment with rating/feature and the power of generating fluent, natural, user review-like text. The main contributions of this paper are summarized as follows\footnote{We have made the code publicly available at: \url{https://github.com/zyrmj0212/CIKM24-MMI-ExR}}:\\
    (1) We identify two key properties crucial for generating useful explanations for recommendations in terms of facilitating users' decision-making: alignment with predicted ratings and item features, which are overlooked in most previous works. And we introduce the mutual information metric to measure the alignment between  explanations and predicted ratings/item features.
    
    (2) We propose a novel MMI framework that features reinforcement learning-based fine-tuning. By customizing reward functions (Mutual Information as the main reward and  KL and Entropy as complementary rewards), the framework can make the pretrained generator align better with the rating or feature while maintaining the ability to mimic user reviews.
    
    (3) We conduct experiments on three public real-world datasets and incorporate different types of backbone models into our MMI framework. The experimental results not only verify the effectiveness of the framework in aligning explanations with predicted ratings and important item features but also demonstrate its capability to strike a good balance between alignment property and similarity with user reviews.
    
    (4) We compare our method with others through human evaluation. The evaluation result further shows the advantage of our method. Additionally, it validates the potential benefits of generating rating-aligned and feature-aligned explanations such as facilitating users' decisions for recommendations from the real users' perspective.

\vspace{-1.0em}
\section{RELATED WORK}
% It's worth mentioning that PEVAE\cite{pevae}  also introduces MI into the DEM (Dependency Maximizing) goal, which enhances the mutual information between the representations of input user-item pairs and corresponding latent variable of VAE to improve the personalization of generated explanations. Compared with our work, PEVAE's MMI goal is a model-intrinsic design tailored for a VAE-based generator, while ours aims at developing a general and model-agnositc fine-tuning framework for different explanation generators. 
% \subsection{Natural language-based Explanation Generation for Recommendation}
 Recent works on generating natural language-based explanations for recommendation can be summarized into two developing lines. The first line adopts more and more advanced model architecture. From RNN-based models \cite{att2seq,nrt,apref2seq,NETE}, to transformer-based \cite{peter,erra} or VAE-based \cite{CVAE,pevae} generators, and now several works \cite{pepler,rexplug,factual-exp,aaai24} have explored the explanation generation ability on LLM. Despite the model architectures of previous works being different, most of them are still trained with the MLE objective may not ensure the alignment properties. The second line endeavors to incorporate rich auxiliary information into the explanation generation. Besides user and item ID, \cite{nrt,att2seq} condition the generation on the rating of the product, \cite{NETE,apref2seq,caml,ucepic} notice the importance of the feature and use pre-defined feature words to guide the generation process. Recently, \cite{factual-exp,erra} have developed retrieval augmented generation to make the generation more personalized and specific. Despite several generators having considered taking predicted rating or item feature as the input, only a few of them \cite{saer,dualpc,erra,NETE} design specific mechanisms to ensure the generated explanation is related to the input rating/feature. Hence, in this work, we focus on designing a model-agnostic fine-tuning framework for existing models to further enhance their ability of generating rating and feature-aligned explanations. 

Our work is similar to two recent works. PEVAE\cite{pevae}  also introduces MI into the optimizing goal, which enhances the mutual information between input user-item pairs and the corresponding latent variable of VAE to improve the personalization of explanations. Compared with our work, PEVAE's MMI goal is a model-intrinsic design tailored for a VAE-based generator, while ours aims at developing a general and model-agnostic fine-tuning framework. LLM2ER-EQR \cite{aaai24} also introduces an RL-based fine-tuning method on GPT-2. However, this paper still treats mimicking user reviews as the main goal by developing specific rewards that measure the similarity between the generated explanation and the review, and it does not examine the impact of the generated explanation on real humans. Comparatively, our designed rewards represent two properties that can bring real benefits to end-users which are further verified through user studies.
% \subsection{Maximizing Mutual Information for Text Generation}
% Maximizing Mutual Information has become an attractive technique for text generation studies \cite{mmi-text-generation-1,mmi-text-generation-2,mmi-text-generation-3,mmi-text-generation-4}. The first work that introduces MMI to text generation is \cite{mmi-text-generation-1}. It treats MMI as a novel decoding method that can help the trained generator produce more diverse, interesting, and appropriate conversational responses. After that, Chawla and Yang \cite{mmi-text-generation-4} introduce MMI to the task of formality style transfer and upgrade the MMI technique as an alternative training objective to the regular maximizing likelihood objective.These successful experiences show the effectiveness of Mutual Information in terms of capturing the correlation between a given context and its corresponding generation. Encouraged by them, this work introduces MI into explanation generation for recommendation and investigates the power of MMI for alleviating the alignment issue for explanation generation.  
% % % Pan et al. \cite{mmi-text-generation-2} and Zhang et al. \cite{mmi-text-generation-3} identify the drawbacks of directly optimizing the MMI problem and design an Adversarial Training framework that adversarially optimizing the mutual information estimator and the text generator. 

\section{Preliminary}
\subsection{Generating Explanation for Recommendation}
We categorize current explanation generation methods into Post-hoc explanation generators and Multi-task Learning models.
\subsubsection{Post-hoc Generation}
Post-hoc explanation generators like \cite{att2seq,apref2seq,compexp,expannet} assume the recommendation has already been made and solely focus on generating an explanation for the given user-item pair $(u,i)$ accompanied by additional attributes such as the rating or a pre-defined feature of the item. They generally adopt a Seq2Seq model architecture that takes some relevant attributes $A=(a_1,a_2,...,a_n)$ of $(u,i)$ as input and use negative log-likelihood (NLL) loss to  maximize the likelihood of generating ground-truth review $e$ conditioned on the given attributes $A$.
\subsubsection{Multi-task Learning}
Multi-task learning models \cite{peter,nrt,dualpc,saer,erra,caml} perform rating prediction and explanation generation simultaneously. Given $(u,i)$, the joint rating-explanation generation task of the models predicts corresponding rating $\hat{r}$ as well as explanation $\hat{e}$. The training objective of the models combines minimizing the mean squared error between $\hat{r}$ and ground-truth rating $r$ and the same NLL loss of ground-truth review $e$ as post-hoc models.\\ \quad
\\
The NLL loss for explanation generation  in both categories is generally defined as:
\begin{equation}
    \begin{aligned}
        L_{e}=-\sum_{w\in e} \log \hat{s}(w)
    \end{aligned}
\end{equation}
where $\hat{s}$ is the predicted word distribution over the vocabulary set.

\subsection{Mutual Information and its Estimation}
\subsubsection{Mutual Information}
Mutual information is an entropy-based measure of dependence between random variables. Given two random variables $X$ and $Y$, the mutual information between them is defined as:

\begin{equation}
    \begin{aligned}
        I(X;Y)=H(X)-H(X|Y)=H(Y)-H(Y|X)=I(Y;X)
    \end{aligned}  
\end{equation}
where $H(X)$ is the Shannon entropy of $X$ and $H(X|Y)$ is the conditional entropy of $X$ given $Y$. As a result, mutual information measures the decrease of the uncertainty in $X$ given $Y$. Intuitively, the higher MI value between  $X$ and $Y$, the stronger dependency there is between $X$ and $Y$ since knowing $Y$ will reduce the uncertainty in $X$.  Such property inspires us to model the alignment degree between explanation and rating or feature with MI and further strengthen it by maximizing MI. 

\subsubsection{Mutual Information Neural Estimation}
\label{sec:mine}
The definition of MI in Eq.(2) can be equivalently expressed as the  KL-divergence between the joint distribution of two variables $X, Y$ and the product of  marginal distribution $P_X$ and $P_Y$:
\begin{equation}
    \begin{aligned}
        I(X;Y)=D_{KL}(\mathbb{P}_{XY}||\mathbb{P}_{X}\otimes \mathbb{P}_{Y} )
    \end{aligned}
\end{equation}
From the above equation, we can see that directly computing MI is intractable because we typically have access to samples but not the underlying distributions \cite{mi-vb,mi-intractable}. Thus, recent works \cite{mi-vb, mine,information-bottleneck,deep-infomax, ba-mi-bound} combine different variational bounds of MI with deep learning to enable differentiable and tractable estimation of mutual information. In this work, we adopt a state-of-the-art method named Mutual Information Neural Estimator (MINE) \cite{mine} to estimate the mutual information between two given variables $X$ and $Y$.
The core idea of MINE is to derive a lower bound of MI  utilizing the following Donsker-Varadhan bound \cite{DV-bound}:
\begin{equation}
    \begin{aligned}
        D_{KL}(P||Q)\geqslant \sup_{T \in F} \mathbb{E}_{P}[T]-\log(\mathbb{E}_{Q}[e^T])
    \end{aligned}
\end{equation}
By combining Eq. (3) and (4) and choosing $F$ to be the family of functions $T_\theta : X \times Y  \to R$ parametrized by a deep neural network with parameters $\theta \in \Theta$, MINE defines following lower bound for true MI:
\begin{equation}
    \begin{aligned}
        I(X;Y)\geqslant I_{\theta }(X;Y)= \sup_{\theta \in \Theta }\mathbb{E}_{\mathbb{P}_{XY}}[T_\theta]-\log (\mathbb{E}_{\mathbb{P}_{X}\otimes \mathbb{P}_{Y}}[e^{T_\theta}])
    \end{aligned}
\end{equation}
In the above equation, $T_{\theta}$ is named as the statistics model in MINE. It takes two variables $X, Y$ as input and outputs a real value. The expectations in the equation are estimated using empirical samples from the joint distribution $\mathbb{P_{XY}}$ and marginal distribution $\mathbb{P_{X}}$ and $\mathbb{P_{Y}}$.
Intuitively, the higher the value of the lower bound is, the more accurate the estimation of true MI is. That means we can treat the lower bound as an optimization goal and adopt a common gradient descent method like SGD to update the statistics model $T_\theta$ iteratively. Once the statistics model $T_{\theta}$ has converged, we can use it to derive an estimated value of MI.

\section{MMI framework}
\begin{figure}
    \centering
    \includegraphics[scale=0.32]{ 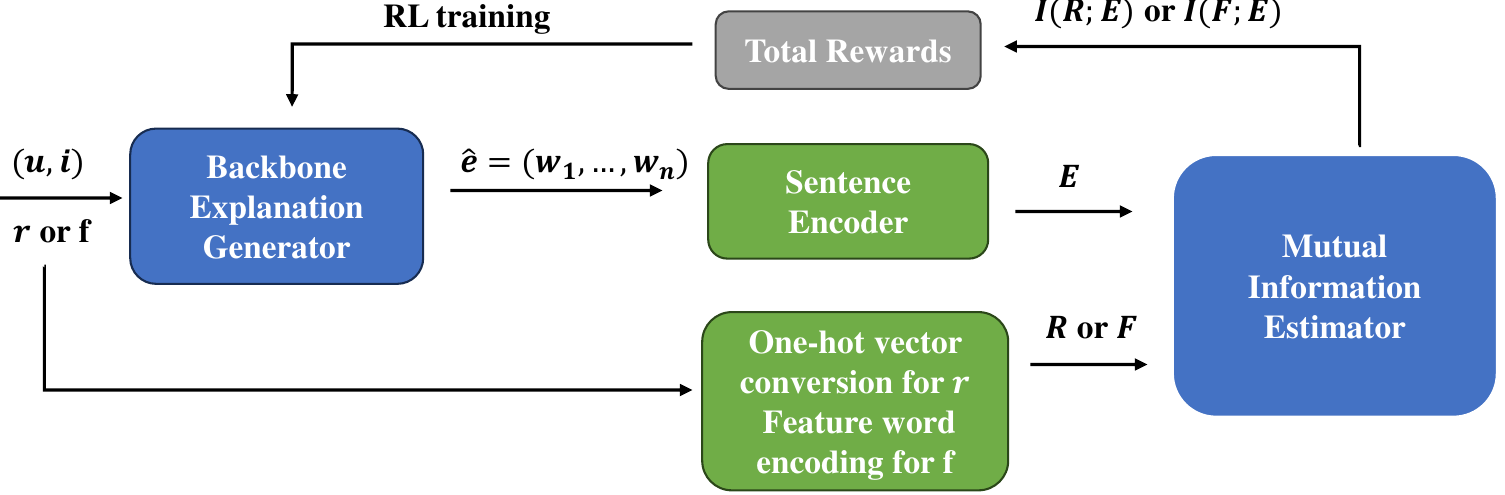}
    \caption{The overview of the MMI framework ( The backbone model in this figure belongs to post-hoc generation models. ) }
    \label{fig:mmi-framework}
    \label{fig:enter-label}
\end{figure}
In our proposed Maximizing Mutual Information (MMI) framework, we start with an arbitrary pre-trained explanation generation model, which we refer to as the  \textbf{backbone} explanation generator. This model has been trained using Maximum Likelihood Estimation (MLE) on review data ( An example loss function is shown in equation (1)), which gives it a strong capability to generate user reviews-like text. And we aim to further enhance its alignment ability through fine-tuning.  The core idea behind this fine-tuning framework is to estimate mutual information (MI) between explanation and rating/feature as a metric to measure the relationship between currently generated explanation and rating/feature. The estimation of MI value requires a complete sentence as the input of a unified sentence encoder, which is obtained by decoding the output log probabilities of the backbone generator, and this decoding process is non-differentiable. Thus it's natural to treat the estimated MI value as a reward and leverage RL to guide the backbone explanation generator in learning better alignment rather than directly treat MMI as an auxiliary loss. Additionally, to maintain the ability of the backbone to simulate user reviews, we also introduce KL and entropy rewards to compensate for the poor text quality incurred by solely optimizing the MI reward.  Figure \ref{fig:mmi-framework} provides an overview of the proposed MMI framework.
\subsection{RL for Fine-tuning Backbone models} 
% Need to specify action, state 
In the RL formulation of explanation generation, the backbone model is considered as an agent and the action is the generation of the word $w_l$ on the next position $t$ based on previous words $w_{1:l-1}$ on position $l-1$. The probability of generation $p_\theta(w_t|w_{1:l-1})$ represents a stochastic policy. We define a customized reward $\pi_{\hat{e}}=\pi_{w_{1:L}} $ at the end of the generated sequence where $L$ is the pre-defined max length of the generated sentence. The optimization goal of the generator $\theta$ is maximizing the expected value of total rewards which induces the following loss function:
\begin{equation}
    \begin{aligned}
        L_{RL}&=- \sum_{\hat{e}}p_{\theta}(\hat{e})\pi(\hat{e})\\
        &=-\sum_{\hat{e}}\prod_{l=1 }^{L-1}p_{\theta}(w_{l+1}|w_{1:l}) \pi (\hat{e})
    \end{aligned}
\end{equation}
We adopt policy gradient to achieve the above optimization goal :
\begin{equation}
    \begin{aligned}
        \nabla_\theta L_{RL}\propto - \pi(\hat{e})\nabla_\theta\log p_\theta(\hat{e})
    \end{aligned}
\end{equation}
The design of the customized reward $\pi_{e}$ is the core of the proposed framework and will be introduced in the next sections.

\subsection{MI Reward for Enhancing Alignment }
To strengthen the alignment of explanation with rating or feature, we propose a Mutual Information (MI) based reward. The MI reward $\pi_{MI}(\hat{e})$ is computed as follows: 1) Use a sentence encoder to transform the generated sample $\hat{e}$ of the explanation generator as a sentence embedding $E$ 2) For alignment with rating task, we convert the 5-level rating score to a 5-dimensional one-hot vector, and for alignment with feature task, we encode the pre-defined item feature word $f$ as a word embedding $F$, 3) We take the concatenation of  $E$ with $R$ or $F$ as the input for the MI estimator, and the output of the estimator will be the MI reward. ( $I(R;E)$ for the task of alignment with rating and  $I(F;E)$ for the task of alignment with feature.). As mentioned in Section \ref{sec:mine}, we adopt MINE \cite{mine} for MI estimation. We denote the statistics model of MINE for $I(R;E)$ as $\theta_{MIR}$ and the one for $I(F;E)$ as $\theta_{MIF}$. According to Eq.(5), we compute MI reward $\pi_{MI}(\hat{e})$ for aligning with rating task as:
\begin{equation}
    \begin{aligned}
        \pi_{MI}(\hat{e})=I_{\theta_{MIR}}(R;E)=\mathbb{E}_{\mathbb{P}_{RE}}[T_{\theta_{MIR}}]-\log (\mathbb{E}_{\mathbb{P}_{R}\otimes \mathbb{P}_{E}}[e^{T_{\theta_{MIR}}}])
    \end{aligned}
\end{equation}
Similarly, MI reward $\pi_{MI}(\hat{e})$ for aligning with feature task is:
\begin{equation}
    \begin{aligned}
        \pi_{MI}(\hat{e})=I_{\theta_{MIF}}(F;E)=\mathbb{E}_{\mathbb{P}_{FE}}[T_{\theta_{MIF}}]-\log (\mathbb{E}_{\mathbb{P}_{F}\otimes \mathbb{P}_{E}}[e^{T_{\theta_{MIF}}}])
    \end{aligned}
\end{equation}
We adopt two strategies to ensure the MI reward model is qualified to give guidance to the generator: 1) we pre-train the MI estimator on the train set of the dataset by treating user review as $E$ and ground truth rating and feature as $R$ and $F$. 2) we alternately update the reward model and the generator in a GAN-like manner to enhance the ability of the reward model in terms of capturing new output samples from the generator. 
\subsection{KL and Entropy Reward for Regularization}
In practice, we observe that without any constraints, the MI-guided fine-tuning process may completely overwrite the original MLE-based policy of the backbone model, leading to a poor text quality of the generated explanation. Meanwhile, as reported in several works \cite{prevent-reward-hacking,reward-hacking-2}, solely optimizing a single reward will incur reward hacking which means the policy exploits loopholes of the reward function and achieves high reward while leading to poor performance and unexpected behaviors. Hence, we introduce the commonly used KL regularization \cite{kl-regularize,prevent-reward-hacking,instructgpt} in RL. The KL reward computes the negative value of KL divergence between the trained new policy and the original policy. In our case, the original policy refers to the pre-trained version of the  backbone model and the new policy is the fine-tuned one, so the KL reward $\pi_{KL}(\hat{e})$ is defined as :
\begin{equation}
    \begin{aligned}
        \pi_{KL}(\hat{e})=-D_{KL}[q(\hat{e})||p_{\theta}(\hat{e})]
    \end{aligned}
\end{equation}
where $q(\hat{e})$  represents the probability of the pre-trained version of the backbone generating the current explanation $\hat{e}$. By maximizing the KL reward, we can reduce the deviation of the fine-tuned model from the pre-trained model and ensure the fine-tuned policy has a safe baseline. 

Additionally, to further increase the diversity of the generation results and facilitate better exploration during RL training, we add Entropy reward as another objective for regularization. The Entropy reward $\pi_{Entropy}$ is computed as :
\begin{equation}
    \begin{aligned}
        \pi_{Entropy}(\hat{e})=H(\hat{e}) =-\sum_{w_l \in \hat{e}}p_{\theta}(w_l|w_{l-1})\log p_{\theta}(w_l|w_{l-1})
    \end{aligned}
\end{equation}

Finally, the total rewards are the weighted summation of the MI,KL and Entropy rewards:
\begin{equation}
    \begin{aligned}
        \pi(\hat{e})=\pi_{MI}(\hat{e}) + \alpha \pi_{KL}(\hat{e}) + \beta \pi_{Entropy}(\hat{e})
    \end{aligned}
\end{equation}
\subsection{Dynamic Weighting Mechanism for Multi-objective Rewards}
According to Eq. (12), the total reward needs to strike a good balance on three different objective rewards. To avoid an exhaustive search of weighting parameters, we propose a dynamic weighting mechanism inspired by the Dynamic Weight Average (DWA) from \cite{dwa} for the three rewards. The dynamic weighting mechanism learns to average rewards weighting over time by considering the rate of change for each reward. Concretely,  the weighting $\gamma $ for reward k $\pi_k$ at time $t$ is defined as:
\begin{equation}
    \begin{aligned}
        \gamma_k(t)=\frac{ Ke^{\frac{h_k(t-1)}{\tau}} } {\sum_{i}e^{\frac{h_i(t-1)}{\tau}}} , 
        h_k(t-1)=\frac{\pi_k(t-2)}{\pi_k(t-1)}
    \end{aligned}
\end{equation}
where $K$ is the number of types of rewards (In our case, $K$=3), $\tau$ is a hyperparameter that controls the softness of the weight distribution. As a result, the dynamic version of the total reward function is \footnote{We only apply the dynamic weighting mechanism on the task of alignment with the feature, since we can easily strike a good balance between different rewards by simply assigning static weights when performing the task of alignment with rating.} :
\begin{equation}
    \begin{aligned}
        \pi_{t}(\hat{e})= \gamma_{MI}(t)\cdot\pi_{MI}(\hat{e}) + \gamma_{KL}(t)\cdot \pi_{KL}(\hat{e}) + \gamma_{Entropy}(t)\cdot \pi_{Entropy}(\hat{e})
    \end{aligned}
\end{equation}
% We only apply the dynamic weighting mechanism on the task of
% alignment with the feature, since we can easily strike a good balance
% between different rewards by simply assigning static weights when
% performing the task of alignment with rating. And we analyze the difference between w./w.o DWA by visualizing the curve of each reward term during the RL process. We observe that w.o. DWA, the RL process is highly sensitive to initial reward weights, with minor tweaks leading to severe fluctuations in the reward curve. Contrastively, adding DWA enables flexible adjustment towards reward weights based on the reward's change ratio, making the RL process more stable and robust.

\subsection{Applying MMI framework on Different Types of Backbone Models}
The general pipeline of applying  MMI framework on backbone models is shown in Figure \ref{fig:mmi-framework}. However, different from the general pipeline, a special adaptation is made when we apply the framework on a multi-task learning model to learn better alignment with rating. This is because, unlike its post-hoc counterpart, the rating $\hat{r}$ is a non-fixed value predicted by the model itself. That means during the fine-tuning process, the rating  $\hat{r}$ will also be updated, which potentially undermines the recommendation performance of the multi-task model. Thus to ensure the recommendation performance will not be affected by the alignment task, we combine the original loss of the backbone model $L_{backbone}$ and the RL objective $L_{RL}$ as the optimization goal of performing alignment with rating task on multi-task learning backbone models:
\begin{equation}
    \begin{aligned}
        L=\lambda L_{RL} +(1-\lambda) L_{Backbone}
    \end{aligned}
\end{equation}

\section{Experiments}
In this section, we conduct extensive experiments to answer the following  four main research questions:
\\ \textbf{RQ1} How does the proposed MMI framework perform in boosting the alignment with predicted ratings and item features?
\\ \textbf{RQ2} To what extent does the MMI framework retain the ability of backbone models to  mimic user reviews?
    % 从具体的实验来说，每个 reward是否起到了应有的作用
\\ \textbf{RQ3} How does each reward, as well as the DWA mechanism benefit the RL fine-tuning process?
% \\ \textbf{RQ4} Can the different adoption of the MMI framework maintain the recommendation performance of multi-task models?
% \\ \textbf{RQ5} Can the effectiveness of the MMI framework in terms of
% alignment with item features be generalized to different feature settings?
% \\ \textbf{RQ6} How does the DWA mechanism benefit the RL fine-tuning process?
\\ \textbf{RQ4} How do real users perceive the MI-enhanced explanation?
\\ \quad

We supplement two more research questions in terms of ablations studies. Due to the space constraint, please refer to an online appendix \footnote{\url{https://github.com/zyrmj0212/CIKM24-MMI-ExR/tree/main/Appendix}} for more details. It’s also worth noting that in this work, we focus on addressing alignment with ratings and alignment with features independently and separately since we intend to closely examine whether the proposed MMI framework can effectively solve each task respectively. However, we do perform a preliminary study to show the potential of our proposed method to simultaneously align generated explanations with both predicted ratings and item features, the corresponding experimental setup and result is also available in our online appendix.
% We supplement two more research questions in terms of ablations studies. Due to the space constraint, please refer to Appendix \ref{rq5-rq6} for more details. 
\subsection{Experimental Setup}
\subsubsection{Datasets}
Experiments are carried out on three real-world datasets from different domains: TripAdvisor (hotels), Yelp (restaurants), and Amazon-MoviesAndTV. We construct the datasets based on the preprocessed version in  \cite{NETE,pepler,peter,erra}, which filters out users with fewer than 5 reviews. The item features in reviews are extracted by Sentires \footnote{https://github.com/evison/Sentires} \cite{sentires}. Additionally, we utilize the Spacy \footnote{https://spacy.io/} toolkit to conduct sentence dependency analysis on each review, removing those where the noun subject is ``I'' or ``We''. This is because such reviews often lack objective descriptions of the items, making them unsuitable to refer to when generating explanations. Finally, we divide the whole dataset into train/validation/test subsets at a ratio of 8:1:1. The details of the datasets are presented in Table \ref{tab:dataset} .

% \tabincell{l}{Examples of  top50 \\ popular features} &  &  & 
\newcommand{\tabincell}[2]{\begin{tabular}{@{}#1@{}}#2\end{tabular}}
\begin{table}[]
\caption{Statistics of datasets}
\label{tab:dataset}
\scalebox{0.9}{\begin{tabular}{l|rrrr}
\hline
                                   & \makecell[c]{TripAdvisor}  & \makecell[c]{ Amazon} &  \makecell[c]{Yelp}\\ \hline
$\#$ users                              & 9,765 & 7,506 & 27,146\\
$\#$ items                             &   6,280& 7,360 & 20,266 \\
$\#$ reviews                           & 269,491 &374,081  &  950,960 \\
$\#$ records per user                   & 27.60 & 49.84  & 35.03 \\
$\#$ records per item                   &  42.91& 50.83 &  46.92 \\
\hline
\end{tabular}}
\end{table}

\subsubsection{Models for Comparison}
%need to add an explanation of the choice of baseline models
We divide existing explanation generation models into two groups to fairly compare our method with other baselines in terms of alignment with predicted ratings and item features, respectively. The baselines in the rating alignment group either take rating as the input of the explanation generator or perform rating prediction and explanation generation simultaneously. The baselines in the feature alignment group all take an item feature as an additional input to the decoder.
\\ 
\textbf{For Rating Alignment:} 
\\
\textbf{NRT} \cite{nrt} jointly model rating prediction and explanation by linearly combining the MSE loss for rating prediction and MLE loss for explanation generation.
\\
 \textbf {Att2Seq} \cite {att2seq} belongs to post hoc generation models. It employs an attribute-to-sequence model architecture to generate an explanation for a product based on the given user, item, and rating of the item. 
% As the rating is only considered as the initial state of the decoder, it cannot ensure the whole generated sentence is indeed related to the input rating.
\\
\textbf{PETER} \cite{peter} integrates the user/item ID information into a transformer-based architecture and introduces a context prediction task to ensure the model generates a unique sentence for each user-item pair. 
% This model also loosely connects rating prediction and explanation generation with sharing part of the model's latent space.
\\
\textbf {PEPLER+MF} \cite{pepler} inputs user and item ID  to a
pre-trained GPT-2 model and perform continuous prompt learning with the MF-based rating prediction task as regularization.
\\
\textbf{DualPC} \cite{dualpc} introduces a duality loss  to closely connect rating prediction and explanation generation tasks. By treating rating predicting as the primal task and explanation generation tasks as the dual tasks, it assumes a well-trained prediction model $\theta_r$ and generation model $\theta_e$ should satisfy the following probabilistic duality:
\begin{equation}
    \begin{aligned}
        p(e)(p(\hat{r}|e;\theta_r))=p(r)p(\hat{e}|r;\theta_e)
    \end{aligned}
\end{equation}
The above equation encourages the generated explanation to align with the ground-truth rating, so there is still a gap in aligning the explanation with the predicted rating. 
\\
\textbf{SAER} \cite{saer} shares a similar motivation of aligning the sentiment of explanation with the predicted rating as our work. It minimizes the difference between the sentiment of current generation and the recommender's prediction through the following loss. The sentiment of the generated explanation is estimated by a pre-trained sentiment regressor $S$.

\begin{equation}
    \begin{aligned}
        L=\sum_{u,i} \mathbb{E}_{P_{\hat{e}|u,i}}[(\hat{r}_{u,i}-f^{S}(\hat{e}))^2]
    \end{aligned}
\end{equation}
 Compared with our MI metric,  SAER's measurement for the relationship between explanation and rating is prone to bias in the sentiment regressor, and it belongs to fitting-based \cite{deep-infomax} metrics, making it less robust and reliable than our proposed MI metric.

% The baselines in this group are NRT \cite{nrt}, Att2Seq \cite{att2seq}, PETER \cite{peter}, PEPLER+MF \cite{pepler}, DualPC \cite{dualpc} and SAER \cite{saer}. 
We apply our MMI framework for rating alignment on \textbf{Att2Seq} and \textbf{PETER}, as they represent two different types of backbones (PETER belongs to multi-task learning models while Att2Seq belongs to post-hoc generation models). We named them as \textbf{Att2Seq + MMI} and \textbf{PETER + MMI}.
\\
\\
\textbf{For Feature Alignment:} 
\\
 \textbf{ApRef2Seq}~\cite{apref2seq} is a Seq2Seq model that encodes historical reviews from users/items and item features as contextual information to
control explanation generation. 
% Although it considers the item feature as the input, it does not guarantee the generated content indeed relates to the given feature.
\\ \textbf{PETER+}~\cite{peter} shares the same transformer-based model architecture as \textbf{PETER}. Different from \textbf{PETER}, it leverages additional feature input as ApRef2Seq to generate more informative explanations. 
% As a result, it shares the same limitation of ApRef2Seq. 
\\  \textbf{PEPLER-D} \cite{pepler} utilizes item features as discrete prompt for pre-trained GPT2. The generator takes the given feature as a prompt word, and generates relevant content revolving around the feature.
\\ \textbf{NETE} \cite{NETE} tailors GRU with a gated fusion unit to incorporate the given feature into the generation. 
\\ \textbf{ERRA}~\cite{erra} inherits the architecture of PETER and features an aspect discriminator loss to encourage the pre-defined item feature to appear in explanation generation.
 %       \begin{equation}
 %        \begin{aligned}
 %            L_a=\frac{1}{T}\sum_{t=1}^{T}(-\tau_a P(w_t|w_{1:t-1}))
 %        \end{aligned}
 %    \end{equation}

 % $\tau_{a}$ is 1 if the generated word at time $t$ is the pre-defined feature,  otherwise, $\tau_{a}$ is 0. 

% The baselines in this group are ApRef2Seq \cite{apref2seq}, PETER+ \cite{peter}, PEPLER-D \cite{pepler}, NETE \cite{NETE} and ERRA \cite{erra}. 

We apply our MMI framework for feature alignment on \textbf{ApRef2Seq} and \textbf{PETER+}. We denoted them as \textbf{ApRef2Seq + MMI} and \textbf{PETER+ + MMI}

\subsubsection{Evaluation Metrics}  

 \quad \\
\textbf{For Rating Alignment:}
\\ \textbf{Normalized Mutual Information }
      We concatenate sentence embeddings of the generated explanations with one-hot vectors representing the predicted rating as input data to train a MINE model. When the model converges, the final value of the lower bound in equation (5)  will be the estimation of $I(R;E)$.  However, due to the predicted rating $R$ (for post-hoc model Att2Seq, we follow previous work by directly using the ground truth rating) of different models are different, we adopt a Normalized version of MI (NMI): \textbf{$\frac{I(R;E)}{H(R)}$} to make the value comparable. NMI ranges from [0,1], the higher the value is, the stronger the alignment of explanation with rating is.
   \\\textbf {Sentiment Accuracy}
    We also conduct sentiment classification tasks on the generated explanations to measure whether the predicted sentiment of the explanation matches the sentiment of the predicted rating. we perform fine-grained (labels are the 1-5 level predicted rating) and coarse-grained (labels are negative, neutral and positive) evaluation, respectively. 
\\ \\
\textbf{For Feature Alignment:}
\\
\textbf{Mutual Information}
    Similar to estimating $I(R;E)$, we concatenate the sentence embeddings of generated explanations with word embeddings of pre-defined item features to train a MINE model. And we can directly compare the estimated \textbf{$I(F;E)$ } of different models since the pre-defined features for all models are the same. 
   \textbf{FMR} Feature Match Ratio \cite{NETE} examine whether the assigned feature $f_{u,i}$ is included in the generated explanation $E_{u,i}$:  $FMR=\frac{1}{N}\sum_{u,i}\mathbb{I}(f_{u,i} \in E_{u,i})$
\\
\textbf{For text similarity with user reviews }
We adopt commonly used metrics for natural language generation: \textbf{BLEU-1} (B-1), \textbf{BLEU-4} (B-4), \textbf{ROUGE-1} (R-1), \textbf{ROUGE-L} (R-L) and \textbf{METEOR} (M) to measure the quality of generated explanation in terms of the similarity with user reviews.
\subsubsection{Implementation details}
%  alignment performance results here
\begin{table*}[]
\centering
\caption{Performance of explanation generation methods in terms of Alignment with Rating}
\label{tab:rating-align-result}
\scalebox{0.8}{
\begin{tabular}{c||c|cc||c|cc||c|cc} 
\hline
                  & \multicolumn{3}{c||}{TripAdvisor}                                                     & \multicolumn{3}{c||}{Amazon}                                                          & \multicolumn{3}{c}{Yelp}                                                               \\ 
\cline{2-10}
\multirow{2}{*}{} & \multirow{2}{*}{$\frac{I(R;E)}{H(R)}$} & \multicolumn{2}{c||}{Sentiment Accuracy~}    & \multirow{2}{*}{$\frac{I(R;E)}{H(R)}$} & \multicolumn{2}{c||}{Sentiment Accuracy~}    & \multirow{2}{*}{$\frac{I(R;E)}{H(R)}$} & \multicolumn{2}{c}{Sentiment Accuracy}        \\ 
\cline{3-4}\cline{6-7}\cline{9-10}
                  &                                        & \multicolumn{1}{l}{5-class} & 3-class        &                                        & \multicolumn{1}{l}{5-class} & 3-class        &                                        & \multicolumn{1}{l}{5-class} & 3-class         \\ 
\hline
NRT               & 0.15                                   & 45.44                       & 87.12          & 0.13                                   & 44.24                       & 75.12          & 0.10                                   & 47.91                       & 69.63           \\
PEPLER+MF         & 0.02                                   & 35.80                       & 65.82          & 0.01                                   & 29.36                       & 59.09          & 0.00                                   & 34.09                       & 61.27           \\
DualPC            & 0.35                                   & 65.84                       & 89.33          & 0.04                                   & 31.08                       & 64.25          &0.43            & 67.32                       & 77.86           \\
SAER              & 0.24                                   & 52.46                       & 87.45          & 0.12                                   & 43.21                       & 73.01          & 0.47                                   & 64.54                       & 82.64           \\ 
\hline
Att2Seq           & 0.22                                   & 47.69                       & 74.66          & 0.27                                   & 49.13                       & 73.35          & 0.31                                   & 51.58                       & 75.11           \\
PETER             & 0.18                                   & 50.68                       & \underline{89.94}  & 0.19                                   & 44.24                       & 75.12          & 0.20                                   & 51.88                       & 71.69           \\ 
\hline\hline
Att2Seq + MMI     & \textbf{0.93}                          & \textbf{76.53}              & 88.89          & \textbf{0.88}                          & \textbf{73.79}              & \underline{89.96}  & \textbf{0.94}                          & \textbf{81.84}              & \textbf{92.50}  \\
PETER + MMI       & \underline{0.44}                           & \underline{70.55}               & \textbf{93.17} & \underline{0.51}                           & \underline{71.39}               & \textbf{96.57} & \underline{0.61}                           & \underline{76.78}               & \underline{88.58}   \\
\hline
\end{tabular}
}

\end{table*}
\begin{table}[]
\centering
\caption{Performance of explanation generation methods in terms of Alignment with Feature}
\label{tab:feature-align-result}
\scalebox{0.8}{
\begin{tabular}{c||cc||cc||cc} 
\hline
                & \multicolumn{2}{c||}{TripAdvisor} & \multicolumn{2}{c||}{Amazon}   & \multicolumn{2}{c}{Yelp}        \\ 
\cline{2-7}
                & $I(F;E)$      & FMR               & $I(F;E)$      & FMR            & $I(F;E)$      & FMR             \\ 
\hline
PEPLER-D        & \underline{2.84}  & \underline{72.98}     & \underline{2.05}  & \textbf{81.50} & \underline{3.06}  & \textbf{88.45}  \\
NETE            & 0.33          & 35.80             & 0.45          & 39.60          & 0.74          & 34.30           \\
ERRA            & 2.29          & 65.84             & 1.42          & 75.71          & 1.55          & 51.48           \\ 
\hline
ApRef2Seq       & 2.19          & 66.79             & 0.92          & 66.24          & 1.23          & 34.31           \\
PETER+          & 1.40          & 55.90             & 0.71          & 52.38          & 1.26          & 46.46           \\ 
\hline\hline
ApRef2Seq + MMI & \textbf{3.12} & \textbf{74.49}    & \textbf{2.15} & \underline{79.45}  & \textbf{3.10} & \underline{85.12}   \\
PETER+ + MMI    & 2.14          & 64.34             & 1.01          & 68.32          & 2.55          & 65.72           \\
\hline
\end{tabular}

}
\end{table}
\quad \\
\textbf{Assignment of item features}
To make the experimental setting more realistic, we assign item features for each model in an estimation manner instead of directly using the features from the reviews in the test set. The estimation method is borrowed from several feature-based explainable recommendation methods \cite{efm,feature-estimation,feature-estimation-2}. First, we select the top 50 popular item features on each dataset \footnote{The reason for using popular features is to ensure the assigned features are all valid. The item features extracted in the  datasets are automatically labeled using Sentires Toolkit. However, in the previous work \cite{efm},  which adopted the same feature-extraction method, the authors asked humans to label the high-quality feature from the extracted feature set. Only around 100 popular features are perceived as qualified features. Moreover, according to our statistics, the frequency of the top 50-top100 features in the three datasets are all below  0.5\%, so we choose Top-50 features as the setting for the main evaluation results.}, then we calculate the user-feature attention vector $x_{ik}$ for each user and the item-feature quality vector $y_{jk}$ for each item as follows, where $t_{ik}$ represents the number of reviews from user $i$ that mentioning feature $k$, $p_{jk}$ represents the number of reviews from item $j$ that mentioning feature $k$ and $s_{jk}$ represents the average sentiment on feature $k$ of item $j$: 
  %   x_{ik}=\left\{ \begin{array}{ c l }
  %  0 & \quad \textrm{if}  \quad t_{ik} =0 \\
  %   1+ (N-1)(\frac{2}{1+e^{-t_{ik}}}-1)     & \quad \textrm{otherwise}
  % \end{array} \right
  
\begin{equation}
\begin{aligned}
      x_{ik}= \begin{cases}
0,  & \text{if $t_{ik}=0$} \\
 1+ (N-1)(\frac{2}{1+e^{-t_{ik}}}-1), & \text{otherwise}
\end{cases}    
\end{aligned}\end{equation}

\begin{equation}
\begin{aligned}
    y_{jk}= \begin{cases}
0,  & \text{if $p_{jk} =0$} \quad\quad \\
 1+ (N-1)\frac{1}{1+e^{-p_{jk}.s_{jk}}},\quad\quad  & \text{otherwise}
\end{cases}    
\end{aligned}\end{equation}

Finally, we assign a feature for each user-item pair according to the dimension with the maximum value in the dot product of the two vectors.
\\

\textbf{Details of the MMI framework}
% describe the training pipeline of the framework, or directly present in the methods section 
% MINE model details (bert encoder, rating and feature variable), feature assignment method ,baseline models parameters ...
We adopt an off-the-shelf BERT model\footnote{https://huggingface.co/bert-base-cased} as the sentence encoder in the framework and the feature word embedding is obtained from the word embedding layer of the sentence encoder. The sentence encoder is fine-tuned on the train set by accomplishing a sentiment classification task. The statistics model used for MI estimation is a three-layer MLP.

\begin{table}[!h]
\caption{Performance of Att2Seq+MMI on Yelp dataset trained with different rewards. Accuracy represents 5-class Sentiment Classification Accuracy. BLEU and Accuracy are percentage values. }
\label{tab:compare-mmi-r-reward}
\scalebox{0.75}{\begin{tabular}{c|ccc|c}
\hline
Reward        & BLEU       & $\frac{I(R;E)}{H(R)}$ & Accuracy & Example output                                        \\ \hline
+MI            & 7.20 0.49  & \textbf{1.00}        & \textbf{85.11}    & \tabincell{c}{Food awesome food amazing \\ amazing delicious delicious} \\ \hline
+MI+KL         & 9.71 0.32  & 0.69        & 77.81    & Amazing food                                          \\ \hline
+MI+KL+Entropy & \textbf{12.91} \textbf{0.69} & 0.94        & 81.84    & \tabincell{c}{The food is amazing and \\the service is top notch  }    \\ \hline
\end{tabular}}

\end{table}
\begin{table}[!h]
\caption{Performance of ApRef2Seq+MMI on Yelp dataset trained with different rewards. BLEU and FMR are percentage values. }
\label{tab:compare-mmi-f-reward}
\scalebox{0.75}{\begin{tabular}{c|ccc|c}
\hline
Reward        & BLEU       & $I(F;E)$ & FMR & Example output                                        \\ \hline
+MI            & 5.63 0.11  & \textbf{3.22}        & \textbf{90.11}    & \tabincell{c}{Sauce sauce good and \\ sauce sauce  great sauce sauce} \\ \hline
+MI+KL         &  8.51 0.05 & 2.69        & 76.79   & The sauce was great                                          \\ \hline
+MI+KL+Entropy & \textbf{12.37 0.60} & 3.10       & 85.12    & \tabincell{c}{The sauce was a little too sweet \\ but the flavor was great }    \\ \hline
\tabincell{c}{+MI+KL+Entropy \\w.o.DWA} & 8.11 0.07 &   2.98   & 80.32    & \tabincell{c}{The sauce was great }    \\ \hline

\end{tabular}}
\end{table}
\subsection{RQ1: Alignment with ratings/features}

\label{sec:main-exp-results}

For RQ1, Table \ref{tab:rating-align-result} and Table \ref{tab:feature-align-result} report the alignment performance of different generation methods. From Table \ref{tab:rating-align-result}, we can conclude that by applying the MMI framework on Att2Seq and PETER, we achieve superior performance in terms of NMI and sentiment accuracy under all settings. Equipped with the MMI framework, the fine-tuned version of Att2Seq and PETER model has gained a stronger alignment ability compared with their pre-trained counterparts, which demonstrates that the MMI framework benefits both multi-task learning model and post-hoc generation model.
% However, we notice that compared with the enhancement of Att2Seq+MMI on Att2Seq, PETER+MMI achieves less improvement over PETER. We speculate that this may caused by the potential conflict between the objective of maximizing mutual information and the multi-task learning goal of PETER. Ablation study results in Section \ref{sec:rq4} verify such conflict.
Besides our  MMI method, SAER and DualPC beat other baseline models on the TripAdvisor and Yelp datasets. That is because they design model-intrinsic mechanisms to relate the explanation generation and rating prediction more closely unlike other models that loosely connect them solely through shared latent space or simply use predicted rating as the initial state of the explanation generator. PEPLER+MF has the worst ability to align with rating, which shows the limitation of prompt-tuning of pre-trained LLM. 

Then, we analyze the results of alignment with features. As shown in Table  \ref{tab:feature-align-result},  our MMI framework enables ApRef2Seq and PETER+ to obtain stronger alignment ability compared with their pre-trained versions and outperforms most baselines in addition to the strongest competitor PEPLER-D. However, we notice that while PEPLER-D can effectively generate sentences containing assigned features, due to its practice of directly using the feature as the prompting word, these sentences are mostly very generic and lack diversity (e.g. ``The food is good.'', ``The service is good.''). Such observation explains the poor performance of PEPLER-D in terms of text generation in Table \ref{tab:text-quality}.

To sum up, the overall evaluation results demonstrate the effectiveness of the proposed MMI framework in enhancing the alignment property of explanation. The improvement based on different backbone models reflects the flexibility and generalizability of the framework to some extent.

\subsection{RQ2: Similarity with user reviews }
Following previous works, we examine all generation methods under traditional NLG metrics to answer RQ2. Based on the results from Table \ref{tab:text-quality}, Att2Seq + MMI, PETER + MMI, ApRef2Seq + MMI and PETER+ + MMI  maintain most of the generation ability of their corresponding backbone models.  That gives RQ2 an affirmative answer that our fine-tuning framework retains most of the knowledge from the pre-training stage to generate fluent, readable, and natural text for end-users. We contribute this advantage to the customized combination of MI, KL, and Entropy reward in the MMI framework, which will be further studied in Section \ref{sec:reward-effect}. Admittedly, in certain settings (e.g. PETER on Yelp dataset),  our method cannot champion in terms of NLG metrics, that is because the ability of our method to simulate reviews is subject to the pre-trained backbone model. When the backbone model performs badly, the MMI fine-tuned version will also not achieve good performance. However, as we have discussed in this paper,  achieving the best NLG metric scores does not necessarily equate to the best quality of explanations in terms of helping users make decisions. And the minor differences in NLG metrics between models might be negligible from real users' perceptions. Therefore, while our method may not excel in NLG metrics, its ability to steer the explanation towards better alignment could still be superior to other approaches in terms of meeting the requirement of explanations in real recommendation scenarios.

\subsection{RQ3: Examine the effect of each reward and DWA mechanism}
\label{sec:reward-effect}
We compare the performance of the method under different reward settings on the Yelp dataset. The results are shown in Table \ref{tab:compare-mmi-r-reward} and Table \ref{tab:compare-mmi-f-reward}. 
% The example output in Table \ref{tab:compare-mmi-r-reward} is an explanation for 5-star rating and one in Table \ref{tab:compare-mmi-r-reward} is an explanation for feature ``sauce'' 
From the results, we observe that solely adopting MI reward attracts the generation process to produce unreadable sentences that contain repetitive keywords which can strengthen the alignment of explanation with rating/feature. That observation echoes our speculation in Section 4.3 that simply adopting MI value as the only reward in RL will incur reward hacking. In our examples, it manifests as the generator finding a shortcut to gain higher reward by the constant addition of words or phrases to the generated sentence that results in high scores for the alignment metric yet the overall quality of the language is severely deteriorated.
Incorporating KL reward forces the generator to remember the knowledge obtained from the pre-trained stage thus improving the text quality to some extent. However it also leads the generator to produce short sentences as the shorter the sequence is, the less discrepancy between the reference model and the generation model will be. Fortunately, entropy reward complements that limitation as it favors longer sentences containing less common words. 

\begin{figure}[b]
    \centering
    \includegraphics[width=0.4\textwidth,height=0.23\textheight]{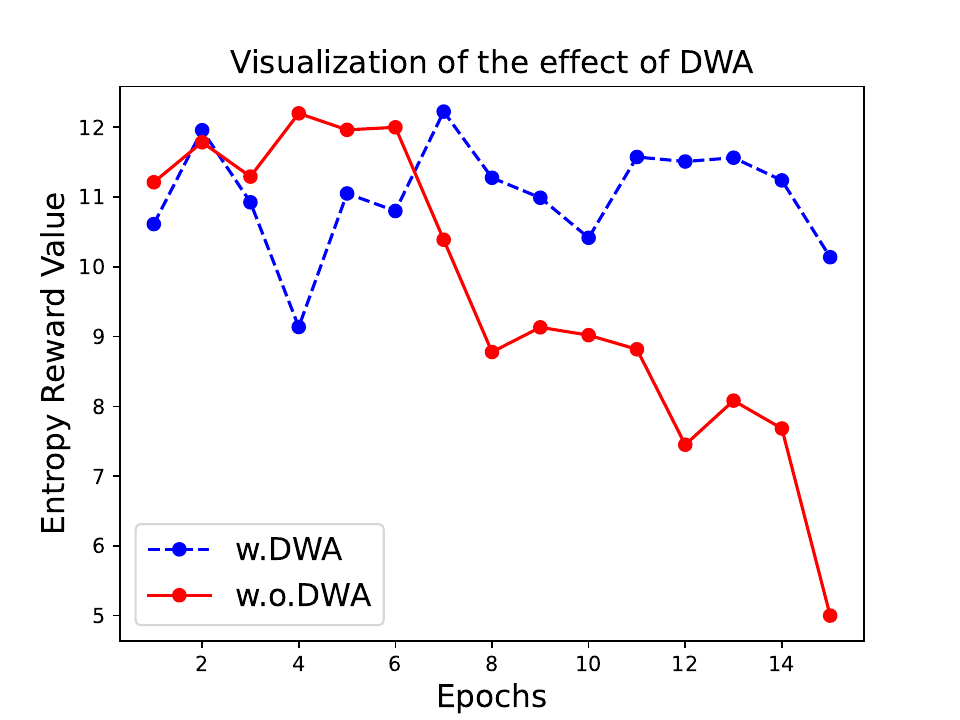}
    \caption{The training curve of Entropy Reward Value under w./w.o. DWA settings. The value of y-axis represents $\gamma_{Entropy}(t).\pi_{Entropy}(\hat{e})$ in equation (14)}
    \label{fig:dwa}
\end{figure}

Meanwhile, we can see the benefits of DWA from  Table \ref{tab:compare-mmi-f-reward}
 and Figure \ref{fig:dwa}. Without DWA, the RL process is highly unstable as we can see severe fluctuation in the entropy reward curve. Such instability will disable the entropy regularizer to control the length of the generated sentence, which corresponds to the poor text quality reflected in Table \ref{tab:compare-mmi-f-reward}.  Contrastively, adding DWA enables flexible adjustment towards reward weights based on the reward's change ratio, making the RL process more stable and robust.
\begin{table*}[]
\centering
\caption{Performance of explanation methods in terms of similarity with user reviews.}
\label{tab:text-quality}
\scalebox{0.8}{\begin{tabular}{c||ccccc||ccccc||ccccc} 
\hline
                & \multicolumn{5}{c||}{TripAdvisor}                                                 & \multicolumn{5}{c||}{Amazon}                                                     & \multicolumn{5}{c}{Yelp}                                                          \\ 
\hline
                & B-1            & B-4           & M              & R-1            & R-L            & B-1            & B-4           & M             & R-1            & R-L            & B-1            & B-4           & M             & R-1            & R-L             \\ 
\hline
\multicolumn{16}{c}{Models for Rating Alignment Comparison}                                                                                                                                                                                                                \\ 
\hline
NRT             & \underline{15.95}  & \textbf{1.12} & \underline{9.87}   & 17.31          & 15.67          & 9.96           & 0.76          & 8.16          & 15.57          & 14.05          & 9.93           & 0.72          & 7.51          & 14.54          & 13.55           \\
PEPLER+MF       & 13.30          & 0.89          & 9.26           & 16.70          & 15.16          & \underline{11.14}  & 0.76          & 8.27          & 14.86          & 13.14          & \underline{11.23}  & \underline{0.76}  & \underline{7.95 } & \textbf{14.89} & 13.81           \\
DualPC          & 13.61          & 0.95          & 9.18           & 16.71          & 15.16          & 10.44          & 0.62          & \textbf{8.75} & \textbf{16.76} & \textbf{15.20} & 10.44          & 0.68          & 7.39          & 14.46          & 13.59           \\
SAER            & 14.83          & 1.02          & 9.66           & 17.21          & \textbf{16.60} & 10.34          & 0.55          & 7.76          & 13.37          & 11.78          & 9.75           & 0.45          & 6.52          & 12.19          & 11.17           \\ 
\hline
Att2Seq         & 15.06          & 0.98          & 9.64           & 17.19          & 15.45          & 10.72          & 0.79          & 8.25          & 15.48          & 13.98          & 10.89          & \textbf{0.81} & 7.76          & \underline{14.87}  & \underline{13.84}   \\
PETER           & 14.82          & \underline{1.07}  & 9.73           & 17.23          & 15.56          & 10.28          & \textbf{0.91} & 8.42          & 15.46          & 13.93          & 8.92           & 0.75          & 7.73          & 15.00          & \textbf{14.02}  \\ 
\hline\hline
Att2Seq + MMI   & \textbf{16.33} & 1.00          & \textbf{10.34} & \underline{17.38}  & 15.42          & \textbf{13.45} & 0.81          & \underline{8.43}  & 15.67          & 13.89          & \textbf{12.91} & 0.69          & \textbf{8.77} & 14.06          & 12.89           \\
PETER + MMI     & 13.97          & 0.94          & 9.68           & \textbf{17.77} & \underline{16.13}  & 10.21          & \underline{0.88}  & 8.22          & 15.42          & 13.79          & 9.11           & 0.66          & 7.90          & 13.97          & 13.77           \\ 
\hline
\multicolumn{16}{c}{Models for Feature Alignment Comparison}                                                                                                                                                                                                               \\ 
\hline
PEPLER-D        & 12.91          & 0.83          & 9.57           & 17.23          & 15.56          & 7.67           & 0.35          & 7.00          & 13.56          & 12.37          & 7.20           & 0.39          & 6.29          & 12.87          & 12.17           \\
NETE            & 13.26          & 1.00          & 9.09           & 16.70          & 15.16          & 10.04          & 0.81          & 8.07          & 15.08          & 13.67          & 10.32          & \textbf{0.71} & 7.51          & 14.22          & 13.27           \\
ERRA            & 15.60          & 1.08          & 9.87           & 17.08          & 15.45          & 12.14          & \textbf{0.95} & \textbf{8.66} & 15.56          & 14.01          & 10.88          & 0.70          & 7.26          & 13.31          & 12.33           \\ 
\hline
ApRef2Seq       & \textbf{16.32} & 1.07          & 10.03          & 17.31          & 15.67          & 13.07          & 0.71          & 8.64          & \textbf{15.87} & \textbf{14.21} & 10.57          & 0.63          & 6.83          & 14.64          & 11.63           \\
PETER+          & 13.93          & \textbf{1.14} & 9.09           & 17.19          & 15.60          & 9.60           & 0.88          & 8.20          & 15.30          & 13.83          & 8.61           & 0.67          & 7.29          & 14.51          & 13.59           \\ 
\hline\hline
ApRef2Seq + MMI & \underline{16.07}  & 0.99          & \textbf{10.12} & \textbf{18.50} & \textbf{17.53} & \textbf{13.90} & 0.39          & 8.49          & 15.48          & 13.37          & \textbf{12.37} & 0.60          & \textbf{7.52} & \textbf{14.70} & \textbf{14.15}  \\
PETER+ + MMI    & 14.30          & 0.95          & 9.32           & 17.27          & 15.55          & 9.45           & 0.76          & 8.13          & 15.02          & 13.43          & 8.32           & 0.44          & 7.15          & 14.35          & 13.93           \\
\hline
\end{tabular}}
\end{table*}

\subsection{RQ4: Human Evaluation}
\begin{table}[]
 \caption{Agreement rate between the model’s predicted rating difference and the users' perceived preference based on the generated explanations.} 
 % (Chi-Square test shows that the agreement rates of different models are significantly different at  $p<0.01$ level.) 
 \label{tab:user_study_a}
\scalebox{0.85}{ \begin{tabular}{r|rrrr}
\hline
            & $\Delta r = 1$ & $\Delta r = 2$ &$\Delta r = 3$ & $\Delta r = 4$ \\ \hline
Att2Seq+MMI &  \textbf{67.61}   &   \textbf{76.00}  &   \textbf{93.33}  &  \textbf{93.5}   \\
Att2Seq     &  31.43   &    56.41 &   57.83  &   51.35  \\
DualPC      &  49.43   &   71.23  &  68.89   &   72.86  \\
SAER        &   40.28  &    45.57 &    42.68 &  63.01   \\
User Review &    30.00 &    57.89 &   64.00  &    63.41 \\
\hline
\end{tabular}}
\end{table}
We recruit 25 participants and design two tasks based on the Yelp dataset. In the first task, we pair items with different ratings and ask participants to choose the item they perceive as the better one based on the generated explanation. We compare 5 methods: Att2Seq, Att2Seq +MMI, SAER, DualPC, and a reference method that directly treats the corresponding user review as the explanation. Each participant is required to annotate 60 records, consisting of 12 records from each explanation method. The comparison results are presented in Table \ref{tab:user_study_a} and grouped by the difference value between the two items' predicted ratings. Att2Seq + MMI archives the best agreement rate under all settings, which indicates that rating-aligned explanations can help users better understand the predicted rating and tell the difference between items. Meanwhile, the relatively poor performance of user reviews highlights the limitations of treating user reviews as ground truth for explanation generation.

\begin{table}[]
 \caption{Human  evaluation of explanations in terms of Informativeness, Relevance, and Satisfaction. }
 % The ANOVA shows that the informativeness, relevance, and satisfaction scores of different models are significantly different at $p<0.01$ level, respectively.  
    \label{tab:user_study_b}
\scalebox{0.85}{ \begin{tabular}{c|ccc}
\hline
            & Informativeness  & Relevance & Satisfaction  \\ \hline
ApRef2Seq+MMI &  \textbf{3.91}   &   \textbf{4.05}  &   \textbf{3.91}    \\
ApRef2Seq    &  3.20   &   2.88 &   3.40   \\
PEPLER-D     & 3.03  &  3.72 & 3.23   \\
ERRA        &  3.51 &   3.40 &   3.50   \\

\hline
\end{tabular}}
\end{table}

In the second task, we sample user-item pairs from the dataset and collect assigned feautes and generated explanations by ApRef2Seq, ApRef2Seq+MMI, ERRA, PEPLER-D. We ask participants to annotate explanations in terms of \textbf{Informativeness \cite{ucepic}} (The generated explanation contains specific information, instead of vague descriptions only.), 
\textbf{Relevance} (The details in the generated explanation are consistent and relevant to the assigned feature of the business.)
and \textbf{Satisfaction} (The generated explanation makes the use of the recommender system fun.). We assign 25 records for each participant and ensure each record has been annotated by at least 3 participants. The annotation results are shown in Table \ref{tab:user_study_b}. 
ApRef2Seq+MMI outperforms other methods in all dimensions, which solidifies the importance of generating feature-aligned explanations. The low informativeness score of PEPLER-D echoes our observation in Section 5.3 that some sentences generated by PEPLER-D are rather generic, providing few details of an item.

\section{Conclusion}
In this paper, we identify the limitation of current explanation generation for recommendation in terms of alignment with the predicted rating and the item feature. To solve this problem, we propose a novel MMI framework, which takes an arbitrary generation model as the backbone and adopts an RL fine-tuning process to maximize the mutual information between the generated explanation and predicted rating/item feature. Experiments on three datasets demonstrate our MMI framework can effectively enhance the alignment ability of different backbone models meanwhile maintaining their ability to simulate user reviews. User studies further confirm the benefits of MI-enhanced explanations to end-users due to their better alignment property.

For future endeavors, we intend to develop a more sophisticated method that can align the explanation with the rating and feature simultaneously, and dig into the relationship between the two tasks deeply. We are also interested in incorporating more properties that are crucial for user-friendly explanations into our fine-tuning framework. For example, as LLM becomes trendy in the field of explanation generation, it will incur the potential risk of factual hallucinations which may deceive end-users and impair users' trust in the recommendation platform. Thus, designing specific metrics to examine this issue will be a non-trivial research topic.

\section{ACKNOWLEDGEMENTS}
We thank the anonymous reviewers for their insightful comments and suggestions. This research was supported by the Natural Science Foundation of China under grant 61902209, 62377044, U2001212, and Beijing Outstanding Young Scientist Program under grant No.BJJWZYJH012019100020098, Intelligent Social Governance Platform, Major Innovation \& Planning
Interdisciplinary Platform for the "Double-First Class" Initiative, Renmin University of China and Public Computing Cloud,Renmin University of China.

\clearpage

\bibliographystyle{ACM-Reference-Format}
\balance

\end{document}